\documentclass[12pt]{iopart}
\usepackage[bookmarks=true, colorlinks=true, linkcolor=blue, urlcolor=blue, citecolor=blue]{hyperref}

\usepackage{iopams}
\usepackage{graphicx}
\usepackage{makecell}
\usepackage{multirow}

\usepackage{mathrsfs}

\usepackage{color}
\begin{document}

\title[Single Entanglement Connection Architecture]{Single entanglement connection architecture between multi-layer bipartite Hardware Efficient Ansatz}

\author{Shikun Zhang$^1$, Zheng Qin$^1$, Yang Zhou$^{1,2,*}$, Rui Li$^1$,Chunxiao Du$^1$ and Zhisong Xiao$^{1,3}$}

\address{$^1$School of Physics, Beihang University, Beijing 100191, China}
\address{$^2$Research Institute for Frontier Science, Beihang University, Beijing 100191, China}
\address{$^3$School of Instrument Science and Opto-Electronics Engineering, Beijing Information Science and Technology University, Beijing 100192, China}

\ead{yangzhou9103@buaa.edu.cn}
\vspace{10pt}

\begin{abstract}
Variational quantum algorithms (VQAs) are among the most promising algorithms to achieve quantum advantages in the NISQ era. One important challenge in implementing such algorithms is to construct an effective parameterized quantum circuit (also called an ansatz). In this work, we propose a single entanglement connection architecture (SECA) for a bipartite hardware efficient ansatz (HEA) by balancing its expressibility, entangling capability, and trainability. Numerical simulations with a one-dimensional Heisenberg model and quadratic unconstrained binary optimization (QUBO) issues were conducted. Our results indicate the superiority of SECA over the common full entanglement connection architecture (FECA) in terms of computational performance. Furthermore,  combining SECA with gate-cutting technology to construct distributed quantum computation (DQC) can efficiently expand the size of NISQ devices under low overhead. We also demonstrated the effectiveness and scalability of the DQC scheme. Our study is a useful indication for understanding the characteristics associated with an effective training circuit.
\end{abstract}

%
\vspace{2pc}

\section{Introduction}

Quantum computing is expected to achieve significant speedups over classical computing for certain applications, such as quantum simulation \cite{1-2,1-3,1-4}, quantum optimization \cite{1-5,1-6}, and quantum machine learning \cite{1-7,1-8}. However, current quantum devices have serious constraints \cite{2}, including limited numbers of qubits, limited connectivity of the qubits, and noisy processes that limit circuit depth \cite{1,1-1}. We are working in the noisy intermediate-scale quantum (NISQ) era \cite{1-1}. One class of algorithms that is expected to unlock the computational potential of NISQ devices is variational quantum algorithms (VQAs) \cite{0,7,18} as they only require the implementation of shallow circuits and simple measurements. Two of representative VQAs are the variations quantum eigensolver (VQE) which is a hybrid algorithm to approximate the ground state eigenvalues for chemical systems  \cite{Peruzzo2014, 5-20} and the quantum approximate optimization algorithm (QAOA) for finding an approximate solution of an optimization problem \cite{1-5, QAOA2}. Also, VQAs have been introduced for other promising applications such as simulation open quantum system dynamics \cite{Mahdian1, Mahdian2, Mahdian3, Mahdian4, Mahdian2-3-20}, which allows us to have a better understanding of the nonequilibrium dynamics of many-body quantum systems \cite{Mahdian3-3-29, Mahdian3-3-31}. 

VQAs use parametrized quantum circuits (PQCs) to run on a quantum computer and then outsource the parameter optimization to a classical optimizer. One important challenge in implementing such algorithms is to construct an effective PQC (also called an ansatz). Generally, such an ansatz should be highly expressive to generate (pure) states that are well representative of the solution space. Besides, the ansatz should also generate highly entangled states with low-depth circuits to efficiently represent the Hilbert space for tasks such as ground state preparation or data classification and to capture non-trivial correlations in quantum data \cite{5-20, 5,5-8,5-9}. Another important factor that should be considered when constructing an effective ansatz is the trainability. It has been shown that quantum landscapes can exhibit the barren plateau (BP) phenomenon,  which means that the gradient of the cost function vanishes exponentially with the number of qubits \cite{20,16-7,16-8,16-9,16-11}. When this occurs, the exponential number of the measurement shots is required to resolve and determine the cost-minimizing direction, thereby diminishing the quantum advantage. Recent studies have demonstrated a trade-off between expressibility, entangling capacity, and trainability. In particular, Ref.\cite{17} has shown that the more expressive a PQC, the smaller the variance in the cost gradient and hence the flatter the landscape; Ref.\cite{18} has found that high entanglement is a hallmark of low-accuracy training and further pointed out that circuit connectivity, and not simply the overall circuit depth, is an accurate indicator of the barrenness of the training landscape. Thus, to construct an effective ansatz, one should balance expressibility, entangling capability, and trainability sophisticatedly.

\begin{figure}
\centering
  \includegraphics[width=0.6\textwidth]{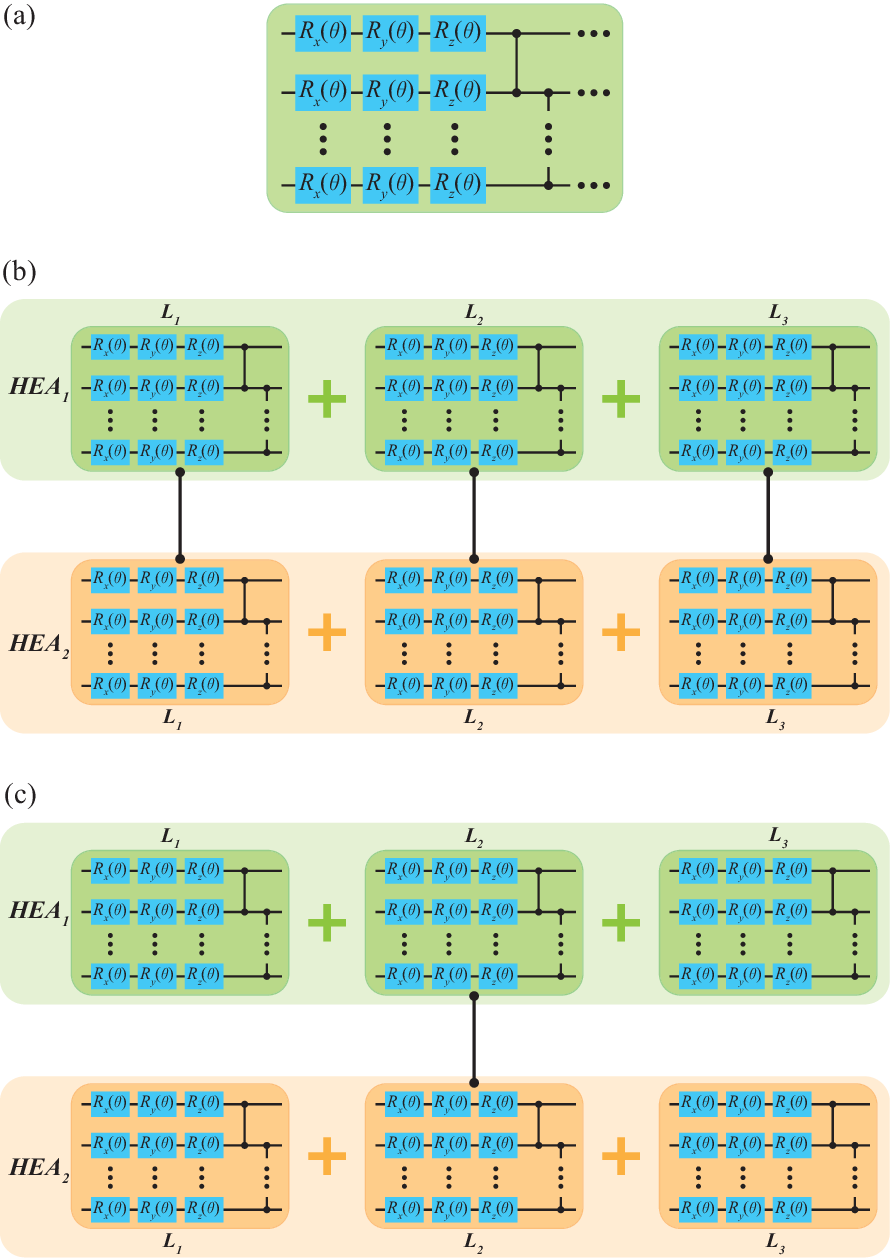}
  \caption{(a) Structure of the HEA. Gates $R_{x}, R_{y}$, and $R_{z}$ are parameterized with $\theta$. (b) Structure of the FECA with $L = 3$. All layers between $HEA_{1}$ and $HEA_{2}$ are connected by $CZ$ gate. (c) Structure of the SECA with $L = 3$. $CZ$ gate connection is only on the middle layer between $HEA_{1}$ and $HEA_{2}$ (i.e., if $L$ is odd, the middle layer is $(L+1)/2$ and if $L$ is even,  the middle layer is $L/2$). For the convenience of calculation, we take $HEA_{1}$ and $HEA_{2}$ with four qubits respectively as an example for analysis.}
  \label{structure}
\end{figure}

In this work, we propose a single entanglement connection architecture (SECA) for a bipartite hardware efficient ansatz (HEA). A standard unit layer of HEA contains single-qubit rotation operations followed by two-qubit entanglement operations (e.g., $CZ$ gate), as shown in Fig.~\ref{structure}(a). In practical applications, the unit layer is generally repeated $L$ times to enhance computational performance \cite{5-20}, which also results in a high degree of circuit connectivity. For simplicity, we divided the original multi-layer HEA into two parts, i.e., $HEA_{1}$ and $HEA_{2}$, delineated in the green and orange boxes respectively, as illustrated in Fig.~\ref{structure}(b). This bipartite division naturally brings about an interesting question of whether it is necessary to perform entanglement connections with $CZ$ gate on every layer between $HEA_{1}$ and $HEA_{2}$ which is called the full entanglement connection architecture (FECA) in our work. In terms of trainability, we should reduce circuit connectivity and suppress the spread of entanglement in PQCs \cite{18}. Thus, we attempt to reduce the entanglement connections (e.g., $CZ$ gate) between $HEA_{1}$ and $HEA_{2}$ to one and call it a single entanglement connection architecture (SECA), as shown in Fig.~\ref{structure}(c).

In fact, there are twofold reasons for proposing the SECA. First, as mentioned above, the trainability of the PQCs is inversely proportional to the number of entanglement connections. By reducing the number of $CZ$ gates ($N_{CZ}$) between $HEA_{1}$ and $HEA_{2}$ to one, the trainability of PQCs can be significantly improved. Second, the roles of entanglement and expressibility in quantum advantages are still unknown. In other words, the amount of entanglement and expressibility required for a PQC to offer an exponential speedup is still an open question. It may not be the case that more entanglement or expressibility is better. One-dimensional Heisenberg model with different coupling strengths and quadratic unconstrained binary optimization (QUBO) problems with different densities of the graph were used to verify the computational performance of SECA. Numerical results have shown that a moderate reduction of entanglement connections does not diminish computational performance, but improves it.

Moreover, there are two additional advantages that make SECA more likely to be deployed on near-term devices. First, entanglement connections were established using non-local two-qubit gates. However, two-qubit gates usually have relatively long latency and high error rates. Our proposed SECA can obtain a better computational performance with fewer two-qubit gates. The quantum computational resources required and the negative effect of quantum noise can be both reduced when experimentally realizing SECA on real quantum hardware. For instance, the realization of quantum hardware based on photonic systems puts strict quality requirements on the entanglement and interference between photons \cite{panjianwei1, panjianwei2}. Accordingly, extensive research efforts have contributed to improving the fidelity and visibility of photonic quantum states to advance photonic quantum hardware development \cite{PPs1, PPs2, PPs3, jichengdanguangzi}. In view of this, implementing SECA on quantum hardware reduces the number of entanglement gates required, which can to some extent alleviate the strict requirements for the fidelity and visibility of photon quantum states. Second, quasi-probabilistic quantum circuit knitting \cite{4} has been proposed to expand the size of NISQ devices. However, a significant issue they face is that the overheads for implementing quantum circuit decomposition grow exponentially with the number of cuts \cite{10,14}. Given the distinctive topological features of SECA, only one cut is required to realize distributed quantum computing (DQC). Our work can provide insights about how to construct an effective ansatz in the NISQ era.

\section{preliminary}\label{sec2}

\subsection{Parameterized Quantum Circuit}

Parameterized quantum circuits (PQCs) are the core component of VQAs. They act as a bridge connecting classical and quantum computing and enable VQAs to simultaneously leverage both classical and quantum computational resources \cite{5}. The output state of the PQCs can be written as
\begin{eqnarray}
  \vert\psi_{\theta}\rangle = U(\boldsymbol\theta)\vert0\rangle^{\otimes{n}},
  \label{equ1}
\end{eqnarray}
which represents applying a parameterized unitary operation $U(\boldsymbol\theta)$ (i.e., ansatz) to an n-qubit reference state $\vert0\rangle^{\otimes n}$. By adjusting the parameters $\boldsymbol\theta$, we can control the final output state $\vert\psi_\theta\rangle$. 

In general, the choice of PQCs, also called ansatz, depends on the tasks to which it will be applied. The evolution process of quantum states may vary significantly among different quantum systems or tasks. Accordingly, there are various ansatz architectures based on the target problems, such as unitary coupled-cluster \cite{5-2, 5-17}, fermionic SWAP network \cite{5-18}, low-depth circuit ansatz \cite{5-19} and so on. In this work, we adopt the so-called hardware-efficient ansatz (HEA) because it is constructed using gates that can be directly executed on near-term quantum hardware. A typical HEA consists of single-qubit gate layers with tunable parameters and two-qubit gate layers to provide entanglement. This can be represented as follows,
\begin{eqnarray}
  U(\boldsymbol\theta) =\prod_{l=1}^{L}U_{l}(\theta_{l})W_{l},
  \label{equ2}
\end{eqnarray}
with 
\begin{eqnarray}
  U_{l}(\theta_{l}) = \bigotimes_{j=1}^{n}R_{\sigma}(\theta_{l}^{j}),
  \label{equ3}
\end{eqnarray}
where $R_{\sigma}(\theta_{l}^{j}) = e^{-i\theta_{l}^{j}\sigma/2}$ with $\sigma \in (\sigma_{x},\sigma_{y},\sigma_{z})$ being one of the Pauli matrices. $l$ represents a specific layer of HEA, whose total number is $L$. $j$ denotes the qubits with the total number being $n$. $W_{l}$ are unparametrized two-qubit gates (e.g., $CZ$ gates). Generally, HEA has a multi-layer architecture and requires the unit layer to be repeated $L$ times to improve their computational performance \cite{5-20}.

\subsection{Trainability}

Trainability is an important factor in the performance of VQAs. Optimization issues such as the barren plateau (BP) have attracted increasing attention \cite{20, 16, 16-7, 16-8, 16-9, 16-11, 16-13, 16-14, 16-15, 16-16}. The so-called BP phenomenon refers to the gradient of a cost function that vanishes exponentially with the number of qubits. To clarify the conditions under which a given ansatz $U(\boldsymbol\theta)$ gives rise to BP, we need to consider a bipartite cut of $U(\boldsymbol\theta)$ with the $k$-th layer as the dividing line to analyze gradient information \cite{20} and rewrite Eq.(\ref{equ2}) as
\begin{equation}
U(\boldsymbol\theta) = U_{\mathcal{L}}(\boldsymbol\theta)U_{\mathcal{R}}(\boldsymbol\theta),
  \label{equ4}
\end{equation}
where 
\begin{eqnarray}
U_{\mathcal{L}}(\boldsymbol\theta) =\prod_{l=k+1}^{L}U_{l}(\theta_{l})W_{l},
\ \ \ U_{\mathcal{R}}(\boldsymbol\theta) = \prod_{l=1}^{k}U_{l}(\theta_{l})W_{l}.
  \label{equ5}
\end{eqnarray}
 For such a $U(\boldsymbol\theta)$ and a certain task with the Hermitian operator $H$, we can define the cost function ($C$) as
\begin{equation}
C = Tr[HU(\boldsymbol\theta)\rho{U(\boldsymbol\theta)^{\dagger}}],
  \label{equ6}
\end{equation}
where $\rho$ is the initial state. By Eq.(\ref{equ4}), we can express the gradient of the cost function with respect to parameter $\theta_{k}$ ($\frac{\partial{C}}{\partial\theta_{k}} := \partial_{\theta_{k}}C$) in a simple form \cite{20}:

\begin{equation}
\partial_{\theta_{k}}C\equiv\frac{\partial{C(\boldsymbol\theta)}}{\partial\theta_{k}}=i\langle0\vert{U^{\dagger}_{\mathcal{R}}}\lbrack{V_{k}},U^{\dagger}_{\mathcal{L}}HU_{\mathcal{L}}\rbrack{U_{\mathcal{R}}}\vert{0}\rangle,
  \label{equ7}
\end{equation}
where $V_{k} = \bigotimes_{j=1}^{n}\sigma_{j}$, is a Hermitian operator.

If $U_{\mathcal{L}}(\boldsymbol\theta)$ and $U_{\mathcal{R}}(\boldsymbol\theta)$ are sufficiently random (or expressive) to form 2-designs, the average value of $\partial_{\theta_{k}}C$ will be zero. If for all $\theta_{k}\in\boldsymbol\theta$, the variance in the cost gradient vanishes exponentially with the number of qubits $n$, i.e.,
\begin{equation}
Var(\partial_{\theta_{k}}C) = \langle(\partial_{\theta_{k}}C)^{2}\rangle \leqslant{O(\frac{1}{\varepsilon^{n}})}\ \ \ for\ all\ \ \ \varepsilon>1,
  \label{equ7}
\end{equation}
the cost function will exhibit BP \cite{20}. According to Chebyshev’s inequality,
\begin{equation}
P(|\partial_{\theta_{k}}C\geqslant\delta|)\leqslant\frac{Var(\partial_{\theta_{k}}C)}{\delta^{2}},
  \label{equ8}
\end{equation}
the probability that $\partial_{\theta_{k}}C$ deviates from its mean value decreases exponentially as the number of qubits $n$ increases, i.e., $\partial_{\theta_{k}}C$ will gradually approach zero. In summary, a sufficiently expressive ansatz exhibits BP, which leads to the disappearance of gradients. Additionally, entanglement in the quantum circuit and noise in NISQ devices can also induce BP. When the BP problem comes up, a large number of iterations are required for the VQA's solution to converge. This is nearly impossible to achieve in practice.

\subsection{Expressibility}

Expressibility refers to the capability of a PQC to produce (pure) states that accurately represent the entire solution space \cite{5}. Numerous studies have characterized the \emph{expressive power} of PQCs \cite{5, exp-prl}. One method to quantify expressibility is to compare the distribution of states obtained through parameter sampling in a PQC with the uniform distribution of states, which corresponds to the ensemble of Haar random states \cite{5}. 

Specifically, we can quantify the expressibility ($Exp$) of a PQC by computing the Kullback-Leibler divergence \cite{5-31} between the fidelity probability distribution of Haar random states and that of PQC states. The corresponding formula is expressed as follows:
\begin{equation}
  Exp := D_{KL}(P_{PQC}(F;\mathbf{\boldsymbol{\theta}}) || P_{Haar}(F)),
  \label{exp}
\end{equation}
$P_{PQC}(F;\boldsymbol{\theta})$ is the estimated fidelity probability distribution of a PQC, where fidelity $F=|\langle\psi_{\boldsymbol\theta}|\psi_{\boldsymbol\phi}\rangle|^2$ is obtained by independently sampling pairs of the parameter vectors of a PQC. $P_{Haar}(F)$ is the fidelity probability density function for the ensemble of Haar random states and can be analytically expressed as \cite{5-30}
\begin{equation}
  P_{Haar}(F) = (N-1)(1-F)^{N-2},
\end{equation}
 where $F$ can take any value between $[0,1]$ and $N$ is the dimension of the Hilbert space. 

In the above definition, expressibility is the amount of information lost if we approximate the distribution of state fidelities generated by a PQC using that of Haar random states. Therefore, a lower value of Exp implies a higher expressibility of a PQC. Thus, the lower the value of $Exp$, the greater the expressibility of a PQC. In addition, because the sample size is finite, to numerically estimate the Kullback-Leibler divergence, it is necessary to discretize both probability distributions by selecting an appropriate binning scheme. In this work, 50 bins were used to compute $Exp$. Although $Exp$ may differ depending on the number of bins, we anticipate that the relative quantitative comparisons among PQCs will remain consistent across observations.

\subsection{Entangling Capability}

In general, the entangling capability of a PQC is linked to its average ability to create entanglements. Most of these measures are based on state entanglement measures. Here, we apply the sampling average of the Meyer-Wallach ($MW$) entanglement measure \cite{5-33} to quantify the entangling capability of a PQC. For a system of $n$ qubits, the $MW$ is defined as follows:
\begin{eqnarray}
  MW(|\psi\rangle) \equiv \frac{4}{n}\sum_{i=1}^{n} \mathcal{D}(\Gamma_{i}(0)|\psi\rangle, \Gamma_{i}(1)|\psi\rangle),
\end{eqnarray}
where $\Gamma_{i}(b)$ is a linear mapping that acts on a computational basis with $b\in\lbrace0, 1\rbrace$, i.e.,
\begin{eqnarray}
\Gamma_{i}(b)|b_{1}\dots{b_{n}}\rangle = \delta_{bb_{i}}|b_{1}\dots{\stackrel{\circleddash}{b}_{i}}\dots{b_{n}}\rangle,
\end{eqnarray}
the symbol $\circleddash$ means to remove the $i$-th qubit. The $\mathcal{D}$ is the generalized distance, which is represented as: 
\begin{equation}
\mathcal{D}(|u\rangle, |v\rangle) = \frac{1}{2}\sum_{i, j}|u_{i}v_{j}-u_{j}v_{i}|^2   
\end{equation}
with $|u\rangle = \sum{u_{i}|i\rangle}$ and  $|v\rangle = \sum{v_{i}|i\rangle}$.
 
As a global measure of multi-particle entanglement for pure states, MW has been widely employed as an effective tool in various quantum information applications \cite{5-33,5-37}, offering insights into entanglement properties. Notably, it is particularly suitable for quantifying entangling capability by evaluating the quantity and diversity of entangled states that PQCs can generate \cite{5}.

In detail, to evaluate the entangling capability ($Ent$) for a PQC, we can approximate $Ent$ by sampling the parameter space and calculating the average Meyer-Wallach measure $(MW)$ of the output states of a PQC, as shown below:
\begin{eqnarray}
  Ent := \frac{1}{|S|}\sum_{\theta_{i}\in{S}}MW(|\psi_{\theta_i}\rangle),
\end{eqnarray}
where $S = \lbrace\theta_i\rbrace$ represents the sets of sampled parameter vectors in the parameter space of a PQC and $|S|$ represents the number of sampled parameter vectors.

\section{Single Entanglement Connection Architecture}\label{sec3}

\begin{figure*}
\centering
\includegraphics[width=1\textwidth]{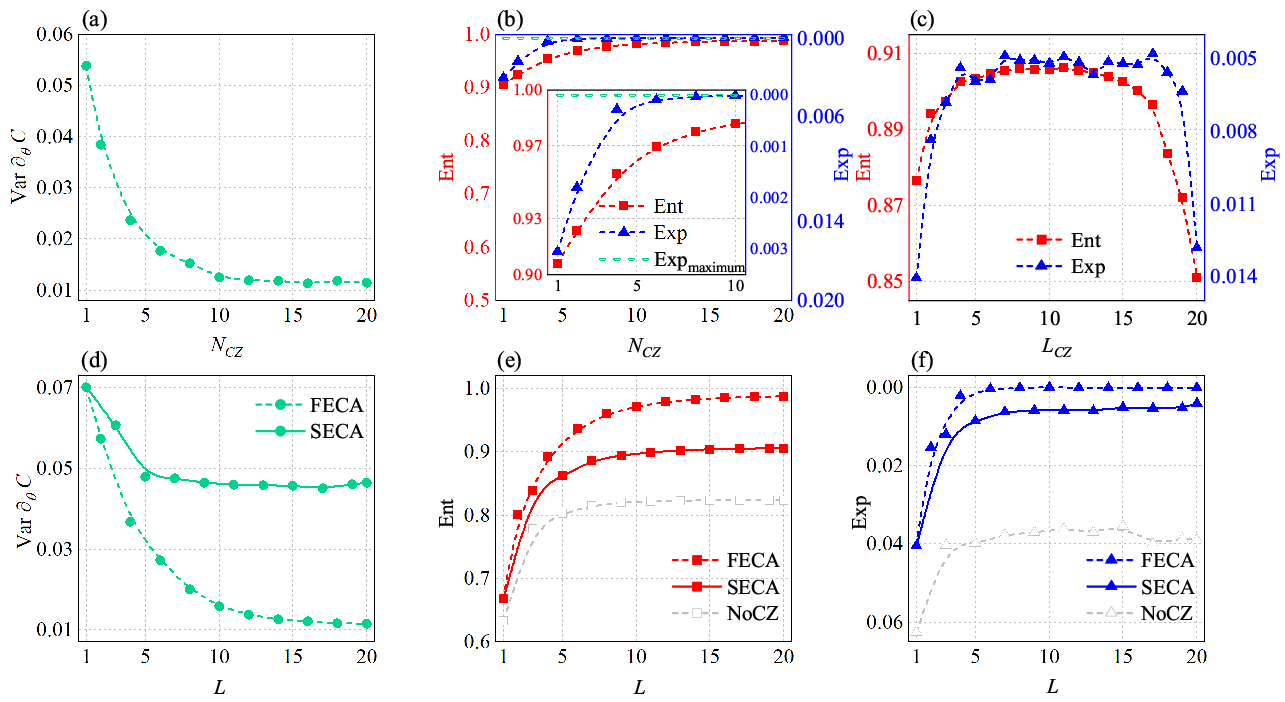}
  \caption{The data analysis graph of trainability, $Ent$ and $Exp$. The y-axis for $Exp$ is inverted to provide a clearer visualization of the increasing and decreasing trends in expressibility. This applies to all FIGs. (a) \& (b) $N_{CZ}$ represents the number of $CZ$ gate between two multi-layer HEAs with $L = 20$. $Var\partial_{\theta}C$ is the variance of the partial derivative of the cost function. (c) $L_{CZ}$ is the layer where the single $CZ$ gate is located on for two multi-layer HEAs with $L = 20$. (d) \& (e) \&(f) $L$ is the number of layers of both FECA and SECA. For SECA, if $L$ is odd, $L_{CZ} = (L+1)/2$ , and if $L$ is even, $L_{CZ}=L/2$.}
  \label{zonghe}
\end{figure*}

\subsection{Construction of the SECA}\label{sec_chara1}

The design of the ansatz architecture is critical to the performance of a variational quantum algorithm. In this section, we explore how to design an efficient ansatz by balancing its expressibility, entangling capability, and trainability. In particular, we study the effect of different numbers of entanglement connections between bipartite HEA and introduce the construction of SECA in detail. 

The most evident difference between FECA and SECA is that they contain a different number of $CZ$ gates. This difference primarily arises from our research methodology of partitioning the original multi-layer HEA into two parts, as shown in Fig.~\ref{structure}(b). The adoption of this research methodology is motivated by the findings in Ref \cite{18}, which infer a proportionality between circuit connectivity, spread of entanglement, and occurrence of BP. This research methodology can effectively control circuit connectivity and the spread of entanglement throughout the entire circuit, concurrently impacting both $Ent$ and $Exp$ of PQCs.
 
 First, we took the number of $CZ$ gates ($N_{CZ}$) between $HEA_{1}$ and $HEA_{2}$ with $L = 20$ as a variable to study the change of $Ent$, $Exp$, and trainability. Notably, a smaller Kullback-Leibler divergence value implies greater expressibility. In order to demonstrate the upward and downward trends of expressibility more clearly, we flipped the y-axis of $Exp$ in Fig.~\ref{zonghe}. In the following text, when we mention a large $Exp$, we no longer refer to the Kullback-Leibler divergence value, but to the actual expressibility, corresponding to the lower value of the Kullback-Leibler divergence and thus the higher positions of points in Fig.~\ref{zonghe}.

Our computational results indicate that as $N_{CZ}$ increases from 1 to 20, trainability (i.e., the variance of the partial derivative of $C$, marked as $Var\partial_{\theta}C$) rapidly decreases and then tends towards a flat landscape, as shown in Fig.~\ref{zonghe}(a). Meanwhile, $Ent$ and $Exp$ both rise rapidly initially. The difference is that $Ent$ continues to rise slowly afterward, but $Exp$ almost saturates to its maximum value ($Exp_{max}$), as shown in the inset of Fig.~\ref{zonghe}(b). Overall, the ranges of variation for both $Ent$ and $Exp$ were not significant.

From these results, we obtain two insights. First, trainability can be significantly improved by reducing $N_{CZ}$. Second, if the maximum $Exp$ is used as the standard for building PQCs, a redundant connection of $CZ$ gates exists. Furthermore, considering that the higher the $Exp$ of the PQCs, the more likely it is to fall into BP \cite{17}, there is no need to perform all entanglement connections between $HEA_{1}$ and $HEA_{2}$. However, determining the optimal $N_{CZ}$ remains a challenge due to that,  so far, we only know that a high $Ent$ or $Exp$ is more inclined to induce BP. Further exploration is required to determine whether or how reducing $Exp$ or $Ent$ changes the cost landscape \cite{17}.

To investigate this issue, we attempted to sacrifice some $Ent$ and $Exp$ to achieve the maximum trainability. $N_{CZ}$ is set to one, which results in some loss of $Ent$ and $Exp$ compared with FECA. The losses are not significant, as shown in Fig.~\ref{zonghe}(b), and they are still sufficient to cover the solution spaces of many problems. This is discussed and proven in Section \ref{cp}. In this case, high trainability and a small search space should significantly improve the optimization performance of the VQAs. This motivated us to further explore this.

Once we have settled on using one $CZ$ gate to link $HEA_{1}$ and $HEA_{2}$, the next step is to determine which layer it should be located on. We label this connection layer as $L_{CZ}$. This is very necessary because different $L_{CZ}$ correspond to different entanglement connection topologies of the PQCs, which are expected to have an impact on both $Ent$ and $Exp$. To identify the optimal $L_{CZ}$, we considered $Ent$ and $Exp$ as functions of $L_{CZ}$. For the bipartite multi-layer HEA with $L = 20$, we compute the values of $Ent$ and $Exp$ when $L_{CZ} = 1$, $L_{CZ} = 2$, all the way up to $L_{CZ} = 20$. It can be seen that both $Ent$ and $Exp$ exhibit a rising-then-falling trend, as shown in Fig.~\ref{zonghe}(c). By selecting the appropriate $L_{CZ}$, we wish to compensate for the loss of $Ent$ and $Exp$ as much as possible. After comprehensive consideration of $Ent$ and $Exp$, we decided to place a single CZ gate in the middle layer, i.e., if $L$ is odd, $L_{CZ} = (L+1)/2$ and if $L$ is even,  $L_{CZ} = L/2$. Up to now, the structure of SECA has been determined. 

Next, we will characterize the SECA in more depth by comparing $Ent$, $Exp$, and trainability with those of the FECA. To map $Ent$, $Exp$, and the trainability of SECA onto a unified scale, we employed the growth rate relative to FECA ($\mathcal{R}$). Take $Ent$ as an example,
\begin{equation}
\mathcal{R}_{Ent} = \frac{\Delta_{Ent}}{FECA_{Ent}}\times100\%, 
\end{equation}
where $\Delta_{Ent} = SECA_{Ent} - FECA_{Ent}$.

Let's first discuss trainability. Fig.~\ref{zonghe}(d) shows that as $L$ increased from 1 to 20, both FECA and SECA exhibited a cost function landscape that first decreased and then flattened, however, the plateau of SECA was much higher than that of FECA. For example, compared to FECA, the maximum $\mathcal{R}_{tra}$ of SECA is approximately $306.7\%$. These results indicate that the trainability of SECA is much higher than that of FECA. 

For $Ent$, as $L$ increases, both FECA and SECA demonstrate an ascending trend with the rate of increase gradually diminishing, indicating that different unit layers have varying effects on $Ent$. Moreover, the $Ent$ gap between SECA and FECA first gradually widened and then stabilized. It must be noted that as $L$ increases, $Ent$ of FECA becomes saturated with very high entanglement values; meanwhile, the circuit connectivity is very high, which increases the probability of BP \cite{18}. In contrast, SECA can limit entanglement to a relatively low level even if $L$ continues to increase. It should be noted that overall, the lost proportion of $Ent$ was not significant. For example, compared with FECA, the maximum $\mathcal{R}_{Ent}$ of SECA is only approximately $-16.6\%$, as shown in Fig.~\ref{zonghe}(e). In addition, to explore the improvement effect of the single CZ gate in SECA for $Ent$, we calculated the case without CZ gate connections (NoCZ) between $HEA_{1}$ and $HEA_{2}$ for comparison, as shown by the gray dashed line in Fig~\ref{zonghe}(e). This result indicates that a single $CZ$ gate in the SECA can significantly improve $Ent$.

Finally, we analyzed and discussed $Exp$. As $L$ increases, the FECA rapidly increases and almost satisfies the 2-designs, see Fig.~\ref{zonghe}(f). As mentioned earlier, this is a hallmark of the BP phenomenon \cite{20}. The SECA shows the same trend, but its saturation value is slightly lower than that of FECA, which means that the probability of BP appearing is smaller than that of FECA. It is worth noting that although $Exp$ of SECA appears almost saturated overall, in fact, $Exp$ always increases slightly, as shown by the solid blue line in Fig.~\ref{zonghe}(f). In addition, we also calculated $Exp$ of NoCZ for comparison. From Fig.~\ref{zonghe}(f), we can see that the single $CZ$ gate in SECA makes a significant contribution to $Exp$, as $Exp$ of SECA is much higher than that of NoCZ. 

It can be seen from the above numerical results that $Ent$ and $Exp$ of SECA are slightly lower than those of FECA. The relative loss is not significant but can effectively prevent $Exp$ and $Ent$ from reaching the upper limit of causing BP. Therefore, SECA has significantly improved trainability compared with FECA. On the whole, in comparison to FECA, the gain of SECA in trainability was much greater than the loss of $Ent$ and $Exp$. There exists a plausible inference that these enhancements significantly improve the computational performance of the SECA.

\subsection{Computational performance of the SECA}\label{cp}

\subsubsection{Evaluation metric}

The most straightforward indicator for evaluating the computational performance of VQE tasks is the expectation value of the energy, which is labeled as $E_{VQE}$. In this section, we introduce a new metric form for VQE tasks, namely the V-score \cite{12}. The V-score allows us to quantify the discrepancy between $E_{VQE}$ and the exact ground state energy, even when the exact ground state energy is unknown. Moreover, the V-score can unify the evaluation of computational performance for different tasks on the same scale. Specifically, the formula of the V-score is as follows:
\begin{equation}
V\mbox{-}score := \frac{N\cdot{E_{var}}}{(E_{VQE} - E_{\infty})^2},
\end{equation}
where $E_{var} = \langle\hat{H}^2\rangle - \langle\hat{H}\rangle^2$. $N$ is the number of degrees of freedom and the constant $E_{\infty}$ plays the role of an energy zero point, compensating for any global energy shift in the Hamiltonian definition. This expression indicates that the closer $E_{VQE}$ is to the exact ground state energy, the closer the V-score is to zero. 

\begin{figure*}
\centering
  \includegraphics[width=1\textwidth]{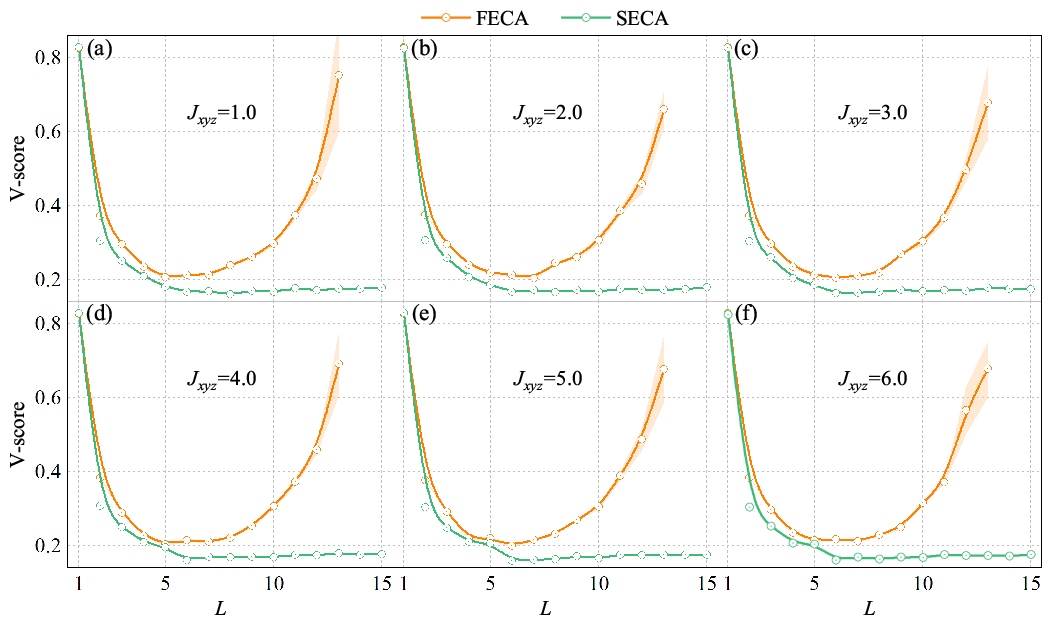}
  \caption{(a)--(f) are the V-score curves of both FECA and SECA as a function of $L$ for the one-dimensional Heisenberg model with coupling strengths $J_{xyz}=1.0, 2.0, 3.0, 4.0, 5.0, 6.0$, respectively. Error bars represent the variance of the results from 100 repeated experiments.}
  \label{Heisenberg}
\end{figure*}

\subsubsection{Heisenberg model}

To verify the computational performance of SECA, we selected two important issues for testing. One of these issues is to solve the ground state of complex quantum systems. The one-dimensional Heisenberg model is chosen as a computational exemplar, which is frequently employed to investigate phase transitions and critical phenomena in magnetic systems and strongly correlated electronic systems. Its Hamiltonian is as follows:
\begin{eqnarray}
  \hat{H} = J\sum_{i=1}^{N}\vec{S}_{i}\cdot{\vec{S}_{i+1}},
\end{eqnarray}
where $S_i$ represents the spin of the $i$th electron and $J$ is the coupling strength. In theory, the entanglement of the ground state can be controlled by changing $J$.  

\subsubsection{Quadratic unconstrained binary optimization}

Another issue is NP-complete and NP-hard combinatorial optimization problems \cite{19,19-6,19-7,19-12,19-18}, one of the most common areas studied in NISQ quantum computing and quantum annealing. Combinatorial optimization problems can be represented as quadratic unconstrained binary optimization (QUBO). QUBO has been applied to solve a wide range of problems such as optimization problems on graphs, resource allocation problems, clustering problems, ordering problems, facility locations problems, and various forms of assignment problems \cite{19}.

Typically, QUBO problems can be conceptualized as an undirected graph comprising $n$ vertices. This undirected graph is interconnected by undirected edges $E(i\Leftrightarrow{j})$, and each $E$ possesses corresponding weights, with $M_{ij}=M_{ji}$. The complexity of QUBO problems can be controlled by \emph{the density of the graph ($D$)},
\begin{equation}
D = \frac{2E}{n(n-1)}
\end{equation}
where $E$ is the number of edges, and $n$ is the number of vertices. The cost function $C$ of QUBO is as follows:
\begin{eqnarray}
C(\vec{x}) = \sum_{i,j}x_{i}M_{ij}x_{j},\ \ \ x_{i}\in{\lbrace0,1\rbrace}.
\label{function}
\end{eqnarray}
where $\vec{x}$ is an $n$ vector of binary variables and $M$ is an $n$-by-$n$ square symmetric matrix of coefficients. The QUBO task involves assigning binary labels (0 or 1) to each vertex and then optimizing cost function $C$ by either maximizing or minimizing it. In practice, QUBO problems are mapped onto an Ising model, where each qubit represents one vertex of the graph. The transformation from a QUBO to an Ising formulation is through the change of variables:
\begin{equation}
x_{i}\longmapsto\frac{1}{2}(1+\sigma_{i}^{z})
\end{equation}
where $\sigma_{i}^{z}$ denotes the Pauli $Z$ matrix acting on the $i$-th qubit. Therefore, the cost function in Eq.(\ref{function}) can be rewritten as follows:
\begin{eqnarray}
f(\sigma_{i}^{z}) = - \frac{1}{2}H + \frac{1}{4}\sum_{i,j}M_{ij},
\end{eqnarray}
with
\begin{eqnarray}
H = \sum_{i<j}J_{ij}\sigma_{i}^{z}\sigma_{j}^{z} +\sum_{i}h_{i}\sigma_{i}^{z}
\end{eqnarray}
where $J_{ij}=-M_{ij}$ is the coupling matrix and $h_{i}=-\sum_{j}M_{ij}$ represents the magnetic fields.

\subsubsection{Results and discussion}

For computing the ground state, we take the one-dimensional Heisenberg model with 8 qubits as an example and choose $L$ as a variable to calculate the V-score values of both FECA and SECA. To verify the computational performance of SECA for various entanglement levels in ground states, we also computed cases with different coupling strengths, i.e., $J_{xyz}=1.0, 2.0, 3.0, 4.0, 5.0, 6.0$, respectively. The corresponding results are shown in Fig.~\ref{Heisenberg}(a-f). All numerical simulation experiments in our work are performed by the \emph{Qiskit} library. \emph{Qiskit} is an open-source quantum computing software development framework developed and maintained by IBM. It aims to provide users with tools and resources for building, running, and analyzing quantum computing experiments. The optimization training steps were set at 5000 iterations, and 100 repetitions of the experiments were performed to enhance the credibility of the experiment. 

\begin{figure}
\centering
  \includegraphics[width=1\textwidth]{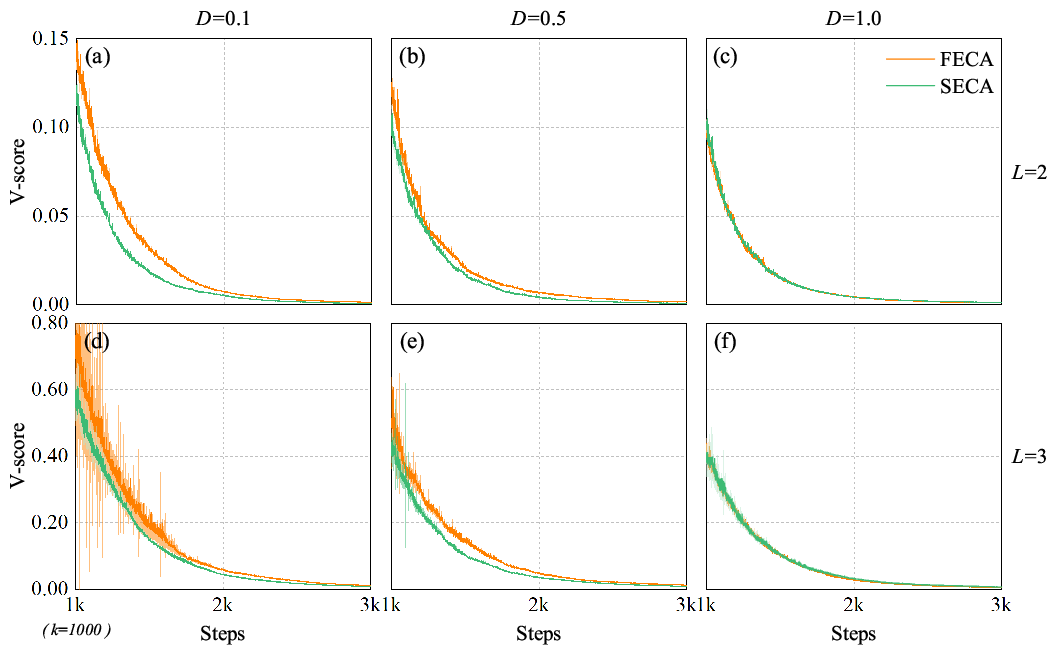}
  \caption{(a)--(f) show the V-score curves of both FECA and SECA as a function of training Steps for QUBO with graph density $D=0.1, D=0.5, D=1.0$, respectively. (a)-(c) corresponds to $L=2$ and (d)-(f) corresponds to $L=3$. Error bars represent the variance of the results from 100 repeated experiments.}
  \label{QUBO}
\end{figure}

It can be seen that as $L$ increases, the computational performance of the FECA exhibits an upward trend followed by a rapid decline with an increase in the error bar. This implies that we can increase $L$ to improve the computational performance, but not the more the better. Regarding the fact that we have fixed the number of optimization training steps, the rapid deterioration of computational performance should be attributed to the rapidly increasing probability of BP. Consequently, for a large $L$, the number of times that the cost function must be evaluated to train the parameters will grow exponentially.

In contrast, the computational performance of SECA rapidly increased with the increase of $L$ and then tended to stabilize without any deterioration. Simultaneously, the error bar was always very small. More importantly, the computational performance of SECA consistently outperformed that of FECA across various coupling strengths ($J_{xyz}$). This demonstrates that although the expressibility and entanglement capability of SECA are weakened, they can still guarantee that the VQE solution is sufficiently close to the exact solution for the problem of interest. Moreover, the computational performance was even enhanced due to the mitigation of the BP phenomenon. 
 
As for QUBO problems, it is always mapped into the Ising model which is relatively simple, and there is no necessity for increasing $L$ to enhance performance. Thus, we set $L=2$ and $L=3$ to calculate the V-score for both FECA and SECA with 12 qubits. For a comprehensive analysis of the optimization process, we treat the training steps as a variable and cap its maximum value at 3000 times. Additionally, some QUBO problems with different $D$ were calculated to verify the computational performance of SECA for different complexities.

As shown in Fig.~\ref{QUBO}(a-f), for both $D=0.1$ and $D=0.5$, SECA exhibits a swifter optimization pace compared to FECA, with slightly superior ultimate computational performance. When $D=1.0$, the computational performances of SECA and FECA are almost the same. It should be noted that QUBO problems with different graph densities $D$ correspond to graphs with different connected vertices. This further verifies that even if we sacrifice part of the expressibility and entanglement capability for the trainability, the space of the generated states is still sufficient to represent the solution space of the target problem. Although both FECA and SECA can address QUBO problems, SECA demonstrates accelerated optimization. Therefore, fewer computational resources are required to achieve commendable performance. For example, fewer times are required for running NISQ devices. In addition, SECA includes a reduced number of entanglement gates, which mitigates the influence of hardware noise on computational performance.

It should be pointed out that during the 100 repeated experiments, we only fixed the value of $D$, this is, the number of connected edges in the graph was fixed, but the connection structure was random. The smaller the value of $D$, the greater the randomness of the graph structure. From the calculation results, it can be seen that compared with FECA, SECA can better adapt to the randomness of the graph structure. This is reflected in SECA's faster optimization speed and smaller error bars of SECA, as shown in Fig.~\ref{QUBO}. We noticed that Fig.~\ref{QUBO}(d) is particularly evident. The reason for the large variance generated by FECA is partly due to the large randomness of the graph structure and partly due to the high $L$.

\section{Distributed quantum computing based on the SECA}

\begin{figure*}
\centering
\includegraphics[width=1\textwidth]{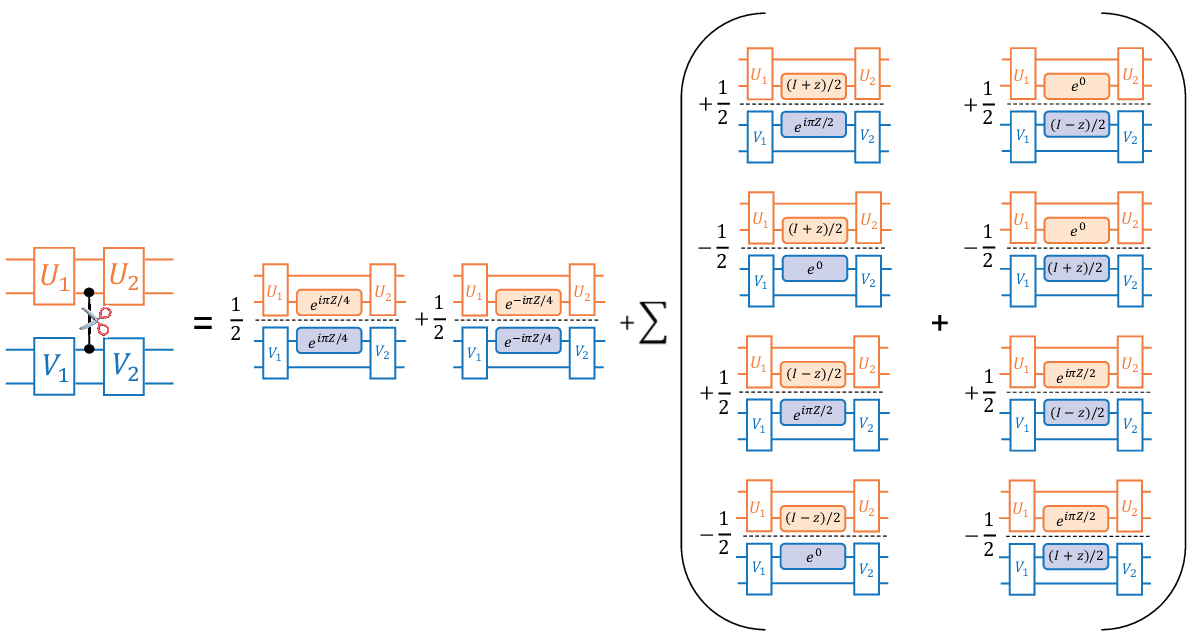}
  \caption{Decomposition of $CZ$ gate into a set of quantum sub-circuits with single-qubit operation.}
  \label{gatecut}
\end{figure*}

Up to now, we have proved that SECA can obtain better stability and computational performance than these of FECA. In this section, we combine SECA with gate cutting technology to construct a distributed quantum computing (DQC) scheme that can help realize large-scale quantum computing on current NISQ devices.

Let's first briefly introduce gate cutting technology. Mitarai and Fuji proposed the construction of a virtual two-qubit gate by sampling single-qubit operations that can simulate the effects of entanglement using classical post-processing and sampling \cite{3}. In addition, it can also be used to simulate a large quantum circuit with small-scale quantum computers. According to their research, any two-qubit gate can be decomposed as follows:
\begin{eqnarray}
  \mathcal{S}&(e^{i\theta{A_{1}\otimes{A_{2}}}}) = cos^{2}\theta\mathcal{S}(I\otimes{I})+sin^{2}\theta\mathcal{S}(A_1\otimes{A_2})\nonumber\\
  &+\frac{1}{8}cos\theta{sin\theta}\sum_{(\alpha_1, \alpha_2) \in \lbrace\pm1\rbrace^2}\alpha_1\alpha_2[\mathcal{S}((I+\alpha_1A_1)\otimes(I+i\alpha_2A_2))\nonumber \\
  &+ \mathcal{S}((I+i\alpha_1A_1)\otimes(I+\alpha_2A_2))],
\end{eqnarray}
where the $\mathcal{S}(U)$ is a superoperator defined by $\mathcal{S}(U) = U\rho{U^\dagger}$, i.e., a unitary operation $U$ onto a quantum state, which is represented by density matrix $\rho$. The operations $\mathcal{S}(I + \alpha{A})$ and $\mathcal{S}(I + i\alpha{A})$ for $\alpha\in\lbrace\pm1\rbrace$ can be implemented by projective measurement on the $A$ basis and single-qubit rotation about the $A$ axis for $A\in\lbrace{X,Y,Z\rbrace}$, respectively. Although circuit cutting technology is a powerful candidate for realizing large-scale quantum computing, a significant issue they face is that both the variance of expectation value computed and the overheads for implementing this quantum circuit decomposition grow exponentially with the number of cuts \cite{10, 14}.

In our proposed SECA, there is only one $CZ$ gate between the bipartite multi-layer HEA. With resort to gate cutting technology, it can be represented by
\begin{equation}
  CZ=e^{i\pi{I\otimes{Z/4}}}e^{i\pi{Z\otimes{I/4}}}e^{-i\pi{Z\otimes{Z/4}}}.
\end{equation}
Then, we can decompose the $CZ$ gate into a sequence of single-qubit operations and present the needed set of quantum sub-circuits for simulating the $CZ$ gate, as shown in Fig.~\ref{gatecut}. This clearly shows that performing one time of gate cutting requires running ten sub-circuits, while the number of sub-circuits executed is $10^L$ for the DQC scheme based on FECA, which requires $L$ gate cutting. Thus, the overhead is reduced from $10^L$ to 10, and it can be concluded that combining the gate-cutting technology with SECA makes the realization of DQC easier while maintaining low overhead. In the NISQ era, this is crucial for expanding the size of the current quantum devices.

Finally, it is necessary to demonstrate the scalability of the DQC scheme. While all the analyses presented in this work are based on the experimental results of 8 or 12 qubits and the range of $L$ is from 1 to 20, the DQC scheme can be extended to accommodate larger numbers of qubits and $L$. Since the primary difference between SECA and FECA is the number of $CZ$ gates employed, we can define the ratio of the number of $CZ$ gates to understand scalability:
\begin{equation}
 R_{CZ} = \frac{S_{CZ}}{S_{FECA}}
 \label{ss}
\end{equation}
where $S_{CZ}$ is the number of $CZ$ gates in the PQCs (e.g., FECA or SECA) and $S_{FECA}$ is the number of $CZ$ gates in the FECA. It is clear that for FECA, $R_{CZ}$ is always equal to 1.

\begin{figure}
\centering
  \includegraphics[width=0.5\textwidth]{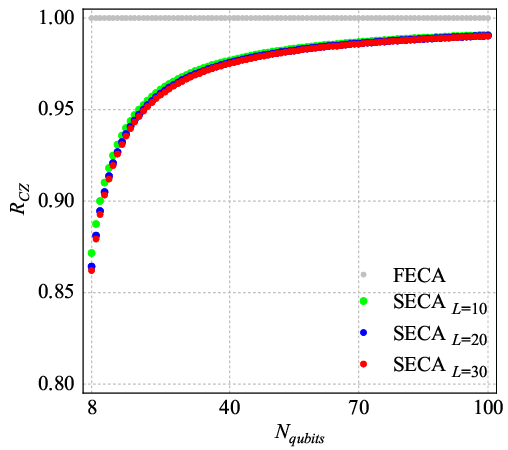}
  \caption{The ratio of the number of $CZ$ gates ($R_{CZ}$) in PQCs compared to FECA, taking the number of qubits ($N_{qubits}$) and $L$ as variables.}
  \label{scala}
\end{figure}

We consider the number of qubits ($N_{qubits}$) as a variable to calculate $R_{CZ}$ for both FECA and SECA, and examine three scenarios with different layers, namely $L=10,20,30$. The numerical results are displayed in Fig.~\ref{scala}. It is evident that as $N_{qubits}$ increases, $R_{CZ}$ also increases gradually at a decreasing rate. Moreover, the different values of $L$ exhibit almost identical trends. This implies that, as $N_{qubits}$ increases, the difference between FECA and SECA gradually diminishes. Although this may have a slight impact on the superior computational performance of SECA, our comprehensive analysis in the preceding text leads us to conclude that the computational performance enhancement by SECA persists with the scaling of both $N_{qubits}$ and $L$ when compared to FECA.

\section{Conclusion}\label{sec5}

In this work, we propose an efficient ansatz architecture for NISQ devices. Two VQE tasks, solving the ground energy of Heisenberg models and addressing QUBO problems, were computed to verify their superiority in computational performance over the common HEA architecture. Different from other studies which only consider one factor for ansatz design, we construct the SECA through a comprehensive analysis of the expressibility, entanglement capability, and trainability of a PQC. Ultimately, we decided to significantly improve trainability by suppressing $Ent$ and $Exp$. The proposed SECA can prevent HEA from reaching the upper bound of the BP generation and alleviate the BP phenomenon. Fortunately, the numerical results demonstrate that weakened $Ent$ and $Exp$ still suffice to solve most questions of interest. Furthermore, combining SECA with gate-cutting technology to construct a distributed quantum computation (DQC) scheme can efficiently expand the size of NISQ devices under low overhead. The scalability of the DQC scheme has also been proven. 

Based on the numerical experiments, we also observed intriguing phenomena that could serve as valuable hints or indications for challenging problems related to the relationships among entangling capacity, expressibility, and computational performance. One of the phenomena is that, as $N_{CZ}$ increases, the entangling capability ($Ent$) increases, but the expressibility ($Exp$) quickly saturates. This may mean that the relationship between $Ent$ and $Exp$ is not simply proportional and that there is a redundant $Ent$ in the multi-layer HEA. An additional observation is that, in comparison to FECA, SECA diminishes the count of CZ gates, resulting in a partial loss of $Ent$ and $Exp$. Nevertheless, this reduction contributes to a substantial enhancement in trainability, consequently leading to a marked improvement in SECA's computational performance. This implies that, for computational performance, there may be the most suitable $Ent$ to achieve optimal performance. Additionally, some recent research shows that both high expressive power and entangling capability can result in a small variance in the cost gradient and hence a flat landscape \cite{17,18,zy3}. Thus, our better computational performance may be attributed to the balance between $Exp$, $Ent$, and trainability. However, this is only a numerical relationship under certain conditions. A more fundamental and general relationship should be established in future research.

Last but not least, our proposed SECA and QDC scheme can be compatible with any mainstream quantum computing platform, such as superconducting quantum computers \cite{chaodao1, chaodao2}, optical quantum computers \cite{panjianwei1, panjianwei2}, and Ion Trap Quantum Computers \cite{lizijing1, lizijing2}, etc., bringing improvements in computing performance and low-cost expansion of computing scale.

\section*{Acknowledgement}
This work was supported by the National Natural Science Foundation of China under Grant No.61975005, the Beijing Academy of Quantum Information Science under Grants No.Y18G28, and the Fundamental Research Funds for the Central Universities under Grants No.YWF-22-L-938.

\section*{References}
\bibliography{ref.bib}

\end{document}